\documentclass[onecolumn,showkeys,pra]{revtex4}
\usepackage{epsf}
\usepackage{amsmath,amsthm,amssymb}
\usepackage{color}
\usepackage{dcolumn}
\usepackage{bm}
\usepackage{graphicx,epsfig}
\usepackage[dvipsnames]{xcolor}

\usepackage{mathtools}

\graphicspath{{fig/}}
\everymath{\displaystyle}

\begin{document}

\title{Production of twisted 
particles in magnetic fields}
 
\author{Liping Zou$^{1}$}
\email{zoulp5@mail.sysu.edu.cn}
\author{Pengming Zhang$^{2}$}
\email{zhangpm5@mail.sysu.edu.cn}
\author{Alexander J. Silenko$^{3,4,5}$}
\email{alsilenko@mail.ru}

\affiliation{$^1$Sino-French Institute of Nuclear Engineering and Technology, Sun Yat-Sen University,
519082 Zhuhai, China}
\affiliation{$^2$School of Physics and Astronomy, Sun Yat-sen University,
519082 Zhuhai, China}
\affiliation{$^3$Bogoliubov Laboratory of Theoretical Physics, Joint
Institute for Nuclear Research,
Dubna 141980, Russia}
\affiliation{$^4$Institute of Modern Physics, Chinese Academy of
Sciences, Lanzhou 730000, China}
\affiliation{$^5$Research Institute for Nuclear Problems, Belarusian
State University, Minsk 220030, Belarus}

\begin {abstract}
The use of a (quasi)uniform magnetic field opens new possibilities for a production of twisted 
particles having orbital angular momenta. Quantum states suitable for a production of charged particles in a uniform magnetic field are determined. 
The particle penetration from a solenoid to vacuum or another solenoid is analyzed in detail. Experiments allowing one successful discoveries of twisted positrons and positroniums are developed. We find a new effect of increasing a uncertainty of phase of the particle rotation with the distance passed by the particle in the solenoid. This effect leads to exciting new possibilities of production of twisted particles in real solenoids without additional devices like particle sources.
\end{abstract}


\maketitle

\section{Introduction}

Twisted (vortex) particles having orbital angular momenta (OAMs) take now an important place in contemporary physics. Twisted photons and electrons have been discovered in 1992 \cite{Allen} and 2010 \cite{UTV}, respectively. The discovery of twisted neutrons \cite{Clark} has been put in question in Ref. \cite{Cappelletti}. Nevertheless, such neutrons have been probably observed \cite{Sarenac}. Vortex beams of atoms and molecules have been obtained in Ref. \cite{HeliumandDimer}.

The wonderful method of production of twisted 
particles in a uniform magnetic field based on the quantum Busch theorem \cite{FloettmannKarlovets,Karlovets2021} have opened new exiting possibilities of a discovery of 
new twisted particles. We confirm main conclusions made in Refs. \cite{FloettmannKarlovets,Karlovets2021} and present a more detailed analysis of physical effects accompanying the production of twisted particles in real solenoids. 

We assume that $\hbar=1,~c=1$ but include $\hbar$ and
$c$ into some formulas when this inclusion clarifies the
problem.


In this Section, we give wave functions and consider main unusual properties of free twisted particle beams. Such beams are now widely used not only in optics but also in electron physics. One of the most important kinds of a twisted beam is a paraxial Laguerre-Gauss (LG) wave beam
\cite{Kogelnik,Siegman,Alda,Pampaloni}. This beam is one of known solutions of the paraxial equation which is obtained in optics and relativistic quantum mechanics in the paraxial approximation ($p_\bot\ll p$):
\begin{equation}
\begin{array}{c}
\left(\nabla^2_\bot+2ik\frac{\partial}{\partial
z}\right)\Psi=0,\qquad
\nabla^2_\bot=\nabla^2-\frac{\partial^2}{\partial z^2}=
\frac{\partial^2}{\partial r^2}+\frac1r\frac{\partial}{\partial
r}+\frac{1}{r^2}\frac{\partial^2}{\partial\phi^2}.
\end{array}
\label{eqp}
\end{equation}
The LG wave beam is given by
\begin{equation}
\begin{array}{c}
\Psi={\cal A}\exp{(i\Phi)},\qquad
{\cal A}=\frac{C_{n\ell}}{w(z)}\left(\frac{\sqrt2r}{w(z)}\right)^{|\ell|}
L_n^{|\ell|}\left(\frac{2r^2}{w^2(z)}\right)\exp{\left(-\frac{r^2}{w^2(z)}\right)},\\
\Phi=\ell\phi+\frac{kr^2}{2R(z)}-\Phi_G(z),\quad
C_{n\ell}=\sqrt{\frac{2n!}{\pi(n+|\ell|)!}},\quad w(z)=w_0\sqrt{1+\frac{4z^2}{k^2w_0^4}},\\
R(z)=z+\frac{k^2w_0^4}{4z},\qquad
\Phi_G(z)=(2n+|\ell|+1)\arctan{\left(\frac{2z}{kw_0^2}\right)},\qquad
\int{\Psi^\dag\Psi r\,dr\,d\phi}=1,
\end{array}\label{eqint}
\end{equation}
where the real functions ${\cal A}$ and $\Phi$ define the amplitude and phase of the beam,
$k$ is the beam wavenumber, $w_0$ is the minimum beam
width, $L_n^{|\ell|}$ is the generalized Laguerre polynomial, $\hbar\ell$ is the OAM, and
$n = 0, 1, 2, \dots$ is the radial quantum number.
This beam
is unlocalized in the longitudinal direction $z$ and transversely 2D-localized.

Some properties of twisted beams are rather unusual.
The recent discovery of the subluminality of structured light beams and its explanation in Ref. \cite{photonPRA}
change perceptions of light quanta. While any photons move with the light velocity, $c$, passing to expectation
values of the momentum and velocity operators (i.e., to semiclassical Einstein quanta) changes their observable properties \cite{photonPRA}. Semiclassical quanta of structured light are subluminal and massive. Their inertial masses coincide with kinematic ones. The quantum-mechanical and semiclassical
descriptions of twisted and other structured electrons lead to similar results. The velocity and the effective mass of the structured photon and electron are quantized \cite{photonPRA}. This effect takes also place for electrons in a uniform magnetic field \cite{arXiv}.

Since twisted electrons have large OAMs, they also possess large magnetic momenta. Additionally, massive particles in  twisted states are characterized by a tensor magnetic polarizability \cite{PhysRevLettLanzhou2019} and a measurable (spectroscopic) electric quadrupole moment \cite{PhysRevLettLanzhou2019,KarlovetsZhevlakov}. In a uniform magnetic field, the effect of a radiative orbital polarization
of a twisted electron beam resulting in a nonzero average projection of the intrinsic OAM on the field direction should take place \cite{ResonanceTwistedElectrons}. We should also mention specific effects at scattering of twisted particles \cite{OAMinteraction,Ivanov:2022jzh}.

In the present study, we analyze the problem of production of particles with OAMs in details. 
We consider a penetration of twisted particle beams produced in a solenoid to vacuum or another solenoid. This penetration is necessary for a discovery of new twisted particles. Unlike previous studies, our analysis is mostly based on gauge-invariant quantum-mechanical and classical equations of motion. Our approach much less uses conservation laws which need taking into account a gauge. We also propose experiments for a discovery of twisted positrons and positroniums.

The paper is organized as follows. 
In Sec. 
\ref{Sect2}, we solve the important problem of quantum states suitable for a production of charged particles in a uniform magnetic field. Canonical and kinetic OAMs of particles produced in real solenoids are considered in Sec. \ref{Sect3}. The key problem of a particle penetration from a solenoid to vacuum or another solenoid is expounded in Sec. \ref{Sect4}. 
In Sec. \ref{QntSt}, we show the existence of new possibilities of production of twisted particles in real solenoids without additional devices like particle sources.
Experiments for a discovery of twisted positrons and positroniums are developed in Sec. \ref{Sect5} and \ref{SecPs}, respectively. The results obtained are summarized in Sec. \ref{SectC}. 

\section{Quantum states suitable for a production of charged particles in a uniform magnetic field}\label{Sect2}

In this Section, we study quantum states suitable for a production of charged particles in a uniform magnetic field. We suppose that $\bm B=B\bm e_z,\, |B|=B$. 

It is instructive to compare the particle production in a uniform magnetic field and the vacuum. 
Hereinafter, basic and nonbasic states denote quantum states eligible and ineligible for such a production. Basic states are used in statistical physics. For particles in the vacuum, a basic state describes 
a plane wave. In this state, the particle momentum $p$ is fixed and the wave function has the simplest form. Certainly, particles can also be created in nonbasic states but only at specific conditions. In 
particular, the production of photons and electrons in states with nonzero OAMs described by the LG beams needs the use of holograms or other appropriate devices. As an example, we can also note an emission of 
twisted photons at a helical motion of charged particles 
\cite{Katoh,Epp,experimental}. Therefore, the quantum states described by the LG beams in vacuum are nonbasic.

For a charged particle in a uniform magnetic field, the simplest quantum states are the Landau ones. For these states, the longitudinal momentum $p_z$ and the squared transversal kinetic momentum $\bm\pi_\bot^2$ are fixed. However, the \emph{general} quantum-mechanical solution has the form of LG beams which are undergoing spatial oscillations and
contains the Landau states as a specific case \cite{arXiv}. We can consider the relativistic particles. 
Relativistic wave functions in the FW representation generalizing the nonrelativistic Landau solution are presented, e.g., in Refs. \cite{paraxialLandau,arXiv}. These wave functions are given by
\begin{equation}
\begin{array}{c}
\Psi={\cal A}\exp{(i\ell\phi)}\exp{(ip_zz)},\qquad \int{\Psi^\dag\Psi rdrd\phi}=1,\\
{\cal A}=\frac{C_{n\ell}}{w_m}\left(\frac{\sqrt2r}{w_m}\right)^{|\ell|}
L_n^{|\ell|}\left(\frac{2r^2}{w^2_m}\right)\exp{\left(-\frac{r^2}{w^2_m}\right)}\eta,\\
C_{n\ell}=\sqrt{\frac{2n!}{\pi(n+|\ell|)!}},\qquad
w_m=\frac{2}{\sqrt{|e|B}}.
\end{array}
\label{Lenergy}
\end{equation}
The spin function $\eta$ is an eigenfunction of the Pauli operator $\sigma_z$:
\begin{equation}\sigma_z\eta^\pm=\pm\eta^\pm,\quad
\eta^+=\left(\begin{array}{c} 1 \\ 0 \end{array}\right),\quad \eta^-=\left(\begin{array}{c} 0 \\ 1 \end{array}\right).
\label{spineta}
\end{equation}
The distinctive feature of the Landau solution is the trivial (exponential) dependence of the electron wave function on $z$. Values of $p_z$ are fixed and $\Psi$ is an eigenfunction of the operator $p_z$.

A transition to the paraxial equation commonly used in the beam theory replaces the exponential factor $\exp{(i\ell\phi)}$ in Eq. (\ref{Lenergy}) with $\exp{(\Phi)}$, where \cite{arXiv,paraxialLandau}
\begin{equation}
\Phi=\ell\phi+\frac{kr^2}{2R(z)}-\Phi_G(z),\qquad \Phi_G(z)=\left(2n+1+|\ell|+\ell+2s_z\right)\frac{z}{z_m},\qquad z_m=\frac{kw_m^2}{2},
\label{eqLGp}
\end{equation}
$\Phi_G(z)$ is the Gouy phase, and $k=p/\hbar$.

Finally, we present the wave functions for a charged particle in a uniform magnetic field obtained in Ref. \cite{arXiv} and having the form of the LG beams. These wave functions are defined by Eq. (\ref{eqint}), but the parameters $w(z),\,R(z)$, and $\Phi_G(z)$ are different:
\begin{equation}
\begin{array}{c}
w(z)=w_0\sqrt{\frac12\left[1+\frac{w_m^4}{w_0^4}-\left(\frac{w_m^4}{w_0^4}-1\right)\cos{\frac{2z}{z_m}}\right]}=w_0\sqrt{\cos^2{\frac{z}{z_m}}+\frac{w_m^4}{w_0^4}\sin^2{\frac{z}{z_m}}},
\\ R(z)=kw_m^2\frac{\cos^2{\frac{z}{z_m}}+\frac{w_m^4}{w_0^4}\sin^2{\frac{z}{z_m}}}{\left(\frac{w_m^4}{w_0^4}-1\right)\sin{\frac{2z}{z_m}}},\\ \Phi_G(z)=(2n+|\ell|+1)\arctan{\left(\frac{w_m^2}{w_0^2}\tan{\frac{z}{z_m}}\right)}+\frac{(\ell+2s_z)z}{z_m}.
\end{array}
\label{general}
\end{equation} In the special case of $w_0=w_m$, the LG beams are equivalent to the Landau beams with the same quantum numbers.

For relativistic particles with a negative charge, the Landau energy levels are defined by (see Ref. \cite{paraxialLandau} and references therein)
\begin{equation}
\begin{array}{c}
E=\sqrt{m^2+p_z^2+(2n+1+|\ell|+\ell+2s_z)|e|B},
\end{array}
\label{eqOAM}
\end{equation} where $n=0,1,2,\dots$ is the radial quantum number and $\ell$ is an eigenvalue of the OAM operator projected on the $z$ axis, $\ell\equiv L_z/\hbar=(\bm r\times\bm p)_z/\hbar$. The symmetric gauge $\bm{A}=\bm B\times\bm r/2,\,A_\phi=Br/2,\,A_r=A_z=0$ is used, where $r$ is the cylindrical coordinate and $r=0$ corresponds to the symmetry axis of particle states. The transverse energies of the Landau levels are defined by the sum of quantum numbers $N=n+(|\ell|+\ell)/2+s_z$. Equation (\ref{eqOAM}) shows that these energies depend on $N$ and the Landau levels are degenerate. For relativistic particles with a positive charge, $\ell$ in this equation should be replaced with $-\ell$. The energy of all states with coinciding $n$ and ${\rm sgn}(e)\ell=0,1,2,\dots$ is the same. Owing to the states with ${\rm sgn}(e)\ell>0$, the degeneracy of the Landau levels is infinite for any $n$. 
We underline that the energy levels (\ref{eqOAM}) are common for the Landau and LG beams.

However, the fact that the Landau states are the simplest ones does not mean that \emph{all} these states are basic. This situation is not extraordinary. The LG beams in a uniform magnetic field, as well as in the vacuum, are solutions of the corresponding quantum-mechanical equations but are not basic. It has been noted in Ref. \cite{paraxialLandau} that the Landau 
states with ${\rm sgn}(e)\ell=0,-1,-2,\dots$ and ${\rm sgn}(e)\ell=1,2,3,\dots$ substantially differ. It can be appropriately proven that the former states are basic and the latter ones are nonbasic. The proof utilizes the definite connection between quantum mechanics and classical physics. It has been shown in Ref. \cite{JINRLett12} with the use of the Wentzel-Kramers-Brillouin method that the classical limit of \emph{relativistic} quantum-mechanical equations is reduced to the replacement of operators in the Hamiltonian and quantum-mechanical equations of motion by the respective classical quantities. Therefore, we can utilize classical electrodynamics for an identification of basic and nonbasic quantum states. 

In the classical picture, there is not any radial motion in a uniform magnetic field. Hereinafter, $\bm\pi=\bm p-e\bm A$ is the kinetic momentum. The classical connection between the kinetic ($\boldsymbol{\mathcal{L}}=\bm r\times\bm\pi$) and canonical ($\bm L=\bm r\times\bm p$) OAMs, as well as the quantum-mechanical one, is given by (see Refs. \cite{FloettmannKarlovets,Wakamatsu})
\begin{equation}\begin{array}{c}
\boldsymbol{\mathcal{L}}=\bm L-\frac e2\bm r\times(\bm B\times\bm r)=\bm L-\frac {e\bm B}{2}r^2,
\end{array}\label{numberd}
\end{equation}
where $r=|\bm r|$ is the radial coordinate in the plane orthogonal to $\bm B$. In the quantum-mechanical picture, $L_z=\hbar\ell$ and $\hbar\ell$ should be close to $-(e/2)B_z r^2$ when $\ell\gg1$. The kinetic OAM, unlike the canonical one, is gauge-invariant (see Ref. \cite{Wakamatsu}). Well-known relativistic classical formulas defining the particle dynamics lead to the following relations:
\begin{equation}
r=|r|=-\frac{\pi_\phi}{eB},\qquad \bm\pi=2\bm p=-2e\bm A,\qquad \mathcal L_z=2L_z=-e Br^2=-\frac{\pi^2_\phi}{eB}.
\label{OAMcl}
\end{equation}

The Landau state radii with respect to the synchrotron motion center are defined by \cite{integral,Li}
\begin{equation}
\langle r^2\rangle=\frac{2(2n+|\ell|+1)}{|e|B},
\label{radii}
\end{equation} Therefore, the connection between the kinetic and canonical OAM operators in the FW representation is given by (see Eq. (\ref{numberd}) where $\bm L$ is an invariant)
\begin{equation}\begin{array}{c}
\langle\mathcal{L}_z\rangle=\ell\hbar-{\rm sgn}(e)(2n+|\ell|+1)\hbar\\=L_z-{\rm sgn}(e)[(2n+1)\hbar+|L_z|],\\ \langle\mathcal{L}_z\rangle-2L_z=-{\rm sgn}(e)[2n+|\ell|+{\rm sgn}(e)\ell+1]\hbar.
\end{array}\label{ckOAM}
\end{equation} Due to the absence of any radial motion, the classical limit of Eq. (\ref{ckOAM}) does not take into account the radial quantum number $n$ and reduces to 
\begin{equation}
\langle\mathcal{L}_z\rangle=L_z-{\rm sgn}(e)|L_z|.
\label{ClLim}
\end{equation} Evidently, the classical limit of the relativistic extension of the Landau result is equivalent to the classical equation (\ref{OAMcl}) when ${\rm sgn}(e)\ell\le0$ and disagrees with it when ${\rm sgn}(e)\ell>0$.

The degeneracy order of the Landau energy levels (\ref{eqOAM}) is equal to infinity owing to the states with ${\rm sgn}(e)\ell>0$. It is generally accepted that a probability of particle production
in any \emph{basic} state is the same for all degenerate states.  If these states were basic, new particles would mostly produced just in them. Since a maximum value of $|\ell|$ is not restricted, created new particles would have the OAMs satisfying the relation ${\rm sgn}(e)\ell>0$ and being very large in absolute value. For so large OAMs, a classical description 
can be fulfilled and its result ($\boldsymbol{\mathcal{L}}=0$) fully contradicts to the classical equation (\ref{OAMcl}). This rigorous result clearly shows that the Landau states with ${\rm sgn}(e)\ell>0$ are not basic for the particle production.  
It can be clearly shown that the states with ${\rm sgn}(e)\ell>0$ are nonbasic not only for $|\ell|\gg1$ but also for $|\ell|\sim1$. The classical limit of relativistic quantum-mechanical equations of spin motion always coincides with the corresponding classical equations even for spins $s=1/2,\,1$ (see Ref. \cite{PRAFW} and references therein). The coincidence takes place due to the well-known perfect correspondence between the commutators of the spin components  $s_i~(i=1,2,3)$ and the related Poisson
brackets in classical mechanics (see Ref. \cite{PRAFW}). The similar correspondence also takes place for the OAM components $L_i~(i=1,2,3)$. In addition, the sum of the spin and OAM is the total angular momentum which forms together with the four-momentum the Poincar\'{e} group (see Ref. \cite{PRAFW} for more details). As a result, the classical limit of relativistic quantum-mechanical equations of OAM motion also should coincide with the corresponding classical equations and \emph{all} states with ${\rm sgn}(e)\ell>0$ are nonbasic.

The new quantum-mechanical results obtained in Ref. \cite{arXiv} make it possible to explain the origin of the states with ${\rm sgn}(e)\ell>0$.
It has been shown \cite{arXiv} that general solutions for charged particles in a uniform magnetic field are LG beams with an oscillating beam width. Oscillations do not take place when the beam waist $w_0$ is equal to the transverse magnetic width of the Landau levels $w_m$ \cite{arXiv}.

In quantum dynamics, turning the total momentum in the plane orthogonal to $\bm e_R$ leads to
the conservation of the radial component of the momentum ($\pi_r=p_r$) which manifests in a permanency of the quantity $$\psi^\dag p_r\psi=-i\hbar\psi^\dag \frac{\partial\psi}{\partial r}.$$ Here $r$ denotes the distance from the axis of symmetry of a quantum state. Amazingly, Eqs. (\ref{eqint}), (\ref{Lenergy}) and the general results obtained in Ref. \cite{arXiv} show that the radial structure of wave functions describing the Landau and LG beams in uniform magnetic fields and the LG beams in the vacuum is the same. We can conclude that crossing the sharp boundary between two areas does not change the width of any Landau beam penetrating from the uniform magnetic field \cite{footnote}. Therefore, the beam width immediately after crossing the boundary remains equal to $w_m$. However, increasing a distance from the boundary leads to widening the beam i.e., increasing $w(z)$ in the vacuum and a circular motion of the beam as a whole (see Eq. (\ref{eqdynll}) below) in a different magnetic field. The corresponding picture in the classical particle physics differs from this quantum-mechanical one by the absence of intrinsic OAMs. For beams penetrated to the vacuum, the classical particle physics predicts a beam momentum spread contrary to non-spreading twisted beams in QM. We can add that multiwave states are appropriately described by the classical \emph{wave} physics.
Equation (\ref{Lenergy}) and Ref. \cite{arXiv} demonstrate that there is also the special case. 
The beam width does not oscillate and a circular motion of the beam does not take place in the field $\widetilde{\bm B}=\widetilde{B}\bm e_z=-B\bm e_z$ because the beam waist in the two fields, $\widetilde{\bm B}$ and $\bm B$, is the same and is equal to $w_m$. In this case, the OAM direction which is unnatural in the field $\bm B~({\rm sgn}(e)\ell>0)$ becomes natural in the field $\widetilde{\bm B}=-\bm B$.
It is instructive to mention that in this case
\begin{equation}\begin{array}{c}
\bm\pi_\bot^2=-\nabla_\bot^2+ieB\frac{\partial}{\partial\phi}+\frac{e^2B^2r^2}{4},\\ \widetilde{\bm\pi}_\bot^2-\bm\pi_\bot^2=e(B+|\widetilde{B}|)\ell+\frac{e^2(\widetilde{B}^2-B^2)r^2}{4}=2eB\ell,\\ \widetilde{p}_z^2-p_z^2=-e(B+|\widetilde{B}|)(\ell+2s_z)-\frac{e^2(\widetilde{B}^2-B^2)r^2}{4}=-2eB(\ell+2s_z).
\end{array}
\label{eqpib}
\end{equation}
The longitudinal momentum remains independent of $r$ and fixed after the penetration into the oppositely directed field only when $\widetilde{\bm B}=-\bm B$ ($\widetilde{B}=-B$). 
This property confirms that the Landau levels with ${\rm sgn}(e)\ell>0$ are nonbasic but can be completed thanks to a penetration of beams with appropriate parameters.

Similar conclusions can be made on a particle production in significantly \emph{nonuniform} magnetic fields. In this case, the classical relation $\mathcal{L}_z=2L_z$ remains unvalid. However, Eq. (\ref{numberd}) shows that $|\mathcal{L}_z|>|L_z|$ and $\boldsymbol{\mathcal{L}}\cdot\bm L>0$ in any case for basic states. In addition, the above-mentioned correspondence between the commutators of the angular momentum and spin components and the related Poisson
brackets in classical mechanics as well as between quantum-mechanical and classical equations of motion leads to a wide agreement between quantum-mechanical and classical results. For quantum states with ${\rm sgn}(e)\ell>0$ in \emph{nonuniform} magnetic fields, $\boldsymbol{\mathcal{L}}\cdot\bm L\le0$ in the classical limit. Therefore, these states are not eligible for a particle production.

We can conclude that the Landau beams with the OAM direction ${\rm sgn}(e)\ell>0$ can be disregarded at a study of the particle production. However, they are not nonphysical and can exist due to a penetration of beams with appropriate parameters.

For real solenoids, $\nabla\cdot \bm B=0$, $(1/R)\partial(RB_R)/(\partial R)=-\partial B_z/(\partial z)$, where $R$ is the distance from the solenoid axis.
To determine the magnetic field near a boundary between the two areas, we can use known formulas given, e.g., in Ref. \cite{FloettmannKarlovets}. We suppose that $R\ll\mathfrak{L}$, where $\mathfrak{L}$ is the solenoid radius. In the needed approximation,
\begin{equation}\begin{array}{c}
B_R(R,\phi,z)=-\frac{R}{2}\frac{\partial B_z}{\partial z}\equiv-\frac{R}{2}B'_z,\qquad B_\phi(R,\phi,z)=0,\\ B_z(R,\phi,z)=B_z(0,\phi,z)-\frac{R^2}{4}\frac{\partial^2 B_z}{\partial z^2}\equiv B_z(0,\phi,z)-\frac{R^2}{4}B''_z.
\end{array}\label{eqelfillll}\end{equation}
The same coordinate system can be used for an antiparallel magnetic field. In this case, reversing the coordinate axis results in $z\rightarrow-z,\,\phi\rightarrow-\phi$.



The exact relativistic FW Hamiltonian is given by
\cite{Energy3,Case,Energy1,JMP}
\begin{equation}
\begin{array}{c}
i\frac{\partial\Psi_{FW}}{\partial t}=\!{\cal H}_{FW}\Psi_{FW},\qquad {\cal H}_{FW}=\!\beta\sqrt{m^2+\bm{\pi}^2-e\bm\Sigma\cdot\bm B},
\end{array}
\label{eq12new}
\end{equation}
where $\beta$ and $\bm\Sigma$ are the Dirac matrices.
This Hamiltonian acts on the bispinor $\Psi_{FW}=
\left(\begin{array}{c} \Phi_{FW} \\ 0 \end{array}\right)$. The lower spinor is nonzero only for states with a negative total energy (virtual states).

The operator equation of motion in the FW representation is similar to the corresponding classical equation and is given by \cite{JMP} (see also Ref. \cite{arXiv} and references therein)
\begin{equation}
\frac{d\bm\pi}{dt}=\beta\frac{e}{4}\left\{\frac1{\epsilon'},\left(\bm\pi\times\bm B-\bm B\times\bm\pi\right)\right\},\qquad \epsilon'=\sqrt{m^2+\bm{\pi}^2-e\bm\Sigma\cdot\bm B},
\label{eqdynal}\end{equation}
where $\bm\Sigma$ is the spin operator and $\beta$ is the Dirac matrix. If the total particle energy is positive and only the upper FW spinor is used, $\beta$ matrix can be omitted:
\begin{equation}
\frac{d\bm\pi}{dt}=\frac{e}{4}\left\{\frac1{\epsilon'},\left(\bm\pi\times\bm B-\bm B\times\bm\pi\right)\right\}.
\label{eqdynll}\end{equation}
The nonzero value of $d\pi_R/(dt)$ is caused only by the centripetal force conditioning a circular motion of a charge in the transversal plane. This centripetal force orthogonal to $\bm\pi_\bot$ rotates the vector $\bm\pi_\bot$ but does not change its absolute value, $\pi_\bot$. 

\section{Canonical and kinetic orbital angular momenta of particles produced in real solenoids}\label{Sect3}

In the most important applications, an external magnetic field is not perfectly uniform but smoothly depends on coordinates and has the axial symmetry. A change of such a field at a distance of the order of $\sqrt{\langle r^2\rangle}$ is often small and can be even neglected. In particular, such a situation takes place in the important case of a particle production in a real solenoid which has been considered in Refs. \cite{FloettmannKarlovets,Karlovets2021}. We will analyze the particle production in such devices in more details. Evidently, the symmetric gauge should be changed as compared with the previous Section. In the general case, the symmetry axis of the external magnetic field ($z$ axis), i.e., the solenoid axes, does not coincide with that of particle states. Let us determine the canonical and kinetic OAMs relative to any point O on the former axis. The particle states relative to the latter axis are defined by the usual quantum-mechanical equations where the magnetic field is equal to the \emph{local} magnetic field $\bm B(\bm R_0)$. Let us choose the plane normal to $\bm B$ and suppose that the point O and the vectors $\bm r$ and $\bm R_0$ belong to this plane. When the radius of a circular motion of the particle is much smaller than the solenoid radius or the considered particle position is far from the solenoid edge, one can use the approximation $\bm B(\bm R)\approx\bm B(\bm R_0)$. In this case, it is convenient to choose the symmetric gauge of the vector potential as follows:
\begin{eqnarray}
\bm A(\bm R)=\bm A_0(\bm R_0)+\bm{\mathcal{A}}, \qquad \bm A_0=\frac12\bm B(\bm R_0)\times\bm R_0, 
\qquad \bm{\mathcal{A}}=\frac12\bm B(\bm R_0)\times\bm r,\qquad \bm R=\bm R_0+\bm r.
\label{eqelfll}\end{eqnarray}  
Evidently, $\bm B(\bm R_0)$ is a constant. When the particle position is far from the solenoid edge, $B''_z$ is relatively small while $B'_z$ can be taken into account. As a result, $B_z\approx B$, $A_{0\phi}\approx R_0B/2$, and $\mathcal{A}_\phi\approx rB/2$.

In this Section, we consider the particle production far from the solenoid edge. 
We can immediately check that $\langle\bm r\rangle=0,\,\langle\bm{\mathcal{A}}\rangle=0,\,\langle\bm \pi_\bot\rangle=0$ and
\begin{equation}\bm R\times\bm A(\bm R)=\frac12 B_z(\bm R_0)(R^2_0+2\bm R_0\cdot\bm r+r^2),\qquad \langle\bm R\times\bm A(\bm R)\rangle=\frac12 B_z(\bm R_0)(R^2_0+\langle r^2\rangle).\label{eqsec}
\end{equation}

Consideration of the canonical and kinetic orbital angular momenta at particle production is straightforward. When only the vector $\bm{\mathcal{A}}$ is taken into account, one can define the \emph{intrinsic} canonical OAM:
\begin{equation}
L_z^{(i)}=(\bm r\times\bm p^{(i)})_z=-e(\bm r\times\bm{\mathcal{A}})_z=-{\rm sgn}(e)\frac{r|\pi_\phi|}{2}.
\label{eqpubcn}
\end{equation}
The total canonical OAM is given by
\begin{equation}
L_z=(\bm R\times\bm p)_z=(\bm R\times\bm\pi)_z+e(\bm R\times\bm{A})_z.
\label{eqpubtc}
\end{equation}
Since $\bm p^{(i)}=\bm\pi+e\bm{\mathcal{A}}$, $\bm\pi=-2e\bm{\mathcal{A}}$, and $\bm R\times\bm A=\bm r\times\bm A_0+\bm r\times\bm{\mathcal{A}}+\bm R_0\times\bm A_0+\bm R_0\times\bm{\mathcal{A}}$, we obtain
$$\bm L-\bm L^{(i)}=\bm R_0\times\bm\pi+e(\bm r\times\bm A_0+\bm R_0\times\bm A_0+\bm R_0\times\bm{\mathcal{A}})=e(\bm r\times\bm A_0+\bm R_0\times\bm A_0-\bm R_0\times\bm{\mathcal{A}})=e\bm R_0\times\bm A_0=\frac{e\bm B}{2}R_0^2.$$
Evidently, the intrinsic and total canonical OAMs are conserved and do not depend on time.

The kinetic OAM is nonconserved and its evolution is defined by that of $\bm R\times\bm A(\bm R)$. Specifically, 
\begin{equation}
\mathcal{L}_z=L_z-\frac e2B_z(\bm R_0)(R^2_0+2\bm R_0\cdot\bm r+r^2)=L_z-\frac e2B_z(\bm R_0)[R^2_0+r^2+2 R_0r\cos{(\omega t+\phi)}],
\label{eqpub}
\end{equation} where $\omega$ is the particle rotation frequency (cyclotron frequency).

Values of $\mathcal{L}_z$ are extremal at $\omega t+\phi=0,\,\pi$. In classical physics, $r=\pi_\bot/|eB_z|$ and Eq. (\ref{eqpub}) takes the form of Eq. (\ref{OAMcl}) at $R_0=0$. For the electron production ($e<0,B_z=B,\mathcal{L}_z\ge0$), we obtain
\begin{equation}
\mathcal{L}_z=L_z+\frac {|e|B}{2}\left[R^2_0+\frac{\pi^2_\bot}{e^2B^2}+2R_0\frac{\pi_\bot}{|e|B}\cos{(\omega t+\phi)}\right].
\label{equel}
\end{equation}

When $R=0$ ($\bm r=-\bm R_0$) at the moment of particle production ($t=0$), $\mathcal{L}_z=(\bm R\times\bm\pi)_z=0$. In the classical picture $L_z=0$ but in the quantum-mechanical one $L_z\neq0$ \cite{FloettmannKarlovets,Karlovets2021} (see also Eqs. (\ref{ckOAM}) and (\ref{ClLim})). In this specific case, the source of twisted particles is very close to the solenoid axis and, in addition, the \emph{average} kinetic OAM $\langle\mathcal{L}_z\rangle$ is defined by $\pi_\bot$, is equal to $L_z+r\pi_\bot$, and does not depend on $R_0$.

We have considered the classical picture of production of twisted particles. The quantum-mechanical picture differs only by the presence of a radial motion.

\section{Particle penetration from a solenoid to vacuum or another solenoid}\label{Sect4}

Laguerre-Gauss beams can penetrate from
the free space into a magnetic field and the other way round 
\cite{arXiv,FloettmannKarlovets,Karlovets2021} and between two solenoids.
For a description of these effects, a penetration of a charged particle
from a (quasi)uniform magnetic field $\bm B$ into the free space or a solenoid with another (quasi)uniform magnetic field $\widetilde{\bm B}$ antiparallel to $\bm B\,\,(\widetilde{\bm B}=\widetilde{B}\bm e_z=-|\widetilde{B}|\bm e_z)$ can be studied in detail.
First, we consider the case when the symmetry axes of the both solenoids coincide with the symmetry axis of particle states. 
We need not suppose the boundary between the magnetic field and the free space or between two magnetic fields with opposite directions ($\widetilde{\bm B}=-|\widetilde{B}|\bm e_z$) to be sharp.

The radial magnetic field defined by Eq. (\ref{eqelfillll}) does not affect $\pi_R$ when the particle crosses the boundary. As a result, the radial component of the momentum $\pi_r=p_r$ conserves. On the contrary, a change of the azimuthal component of the kinetic momentum takes place. In the semiclassical approximation, it is given by
\begin{equation}
\frac{d\pi_\phi}{dt}=\frac{eB_R\pi_z}{\epsilon}=eB_R\frac{dz}{dt}.
\label{emeqp}\end{equation}  We disregard the orbital motion here.
As a result of the integration,
\begin{equation}
\Delta\pi_\phi=\int{\frac{eB_R\pi_z}{\epsilon}dt}=-\frac{eR}{2}\int{\frac{\partial B_z}{\partial z}dz}=\frac{eR(B+|\widetilde{B}|)}{2}.
\label{eqinm}
\end{equation} It follows from Eq. (\ref{eqinm}) that the azimuthal component of the kinetic momentum changes. It can be checked that the azimuthal component of the canonical one remains unchanged. The last property provides for the conservation of the canonical (but not kinetic) OAM.
If the spin-dependent interaction is neglected, we obtain that the total momentum is turned through the definite angle in the plane orthogonal to $\bm e_R$. The result obtained shows that the internal canonical momentum $p_\phi$ conserves immediately after the penetration. Indeed, $\Delta p_\phi=\Delta\pi_\phi+e\Delta A_\phi$ and the change of the vector potential is defined by
\begin{equation}
\Delta\bm A=\frac{\Delta\bm B\times\bm R}{2}=-\frac{R(B+|\widetilde{B}|)}{2}\bm e_\phi.
\label{eninm}
\end{equation} As a result, $\Delta p_\phi=0$.

We can now pass to a more general case when the the symmetry axes of the both solenoids coincide but differ from the symmetry axis of particle states. We underline that all partial waves remain coherent in any magnetic field because their \emph{total} energies are not changed after passing through the boundary. A certain complication takes place for \emph{spin-polarized} particle beams owing to the interaction of magnetic moments of particles with the magnetic field. In this case, penetrating particles are affected by the additional small spin-dependent force proportional to $B'_z$. This force slightly changes the longitudinal momentum and the kinetic energy but does not influence the transversal momentum and the total energy. Any partial wave forming the particle state freely passes through the boundary between the two areas.

Our detailed analysis confirms the main conclusion made in Ref. \cite{FloettmannKarlovets}. Charged particles produced in a magnetic field and penetrating from it are twisted. 
In classical particle physics, a particle motion is characterized by any continuous trajectory. The relativistic quantum mechanics (QM) in the FW representation and the paraxial QM use the Schr\"{o}dinger quantum-mechanical picture. One of fundamental properties of Schr\"{o}dinger QM is a continuity of wave functions. As a result, the particle cannot be in a plane wave state and in one of Landau states after a penetration from a solenoid to vacuum or another solenoid, respectively. The case of $\widetilde{\bm B}=-\bm B$ is an exclusion. We should also take into account that crossing the boundary conserves the total canonical OAM and the quantity $\bm{\pi}^2-e\bm\Sigma\cdot\bm B$, where $\bm\Sigma$ is the spin operator (see Ref. \cite{arXiv} and references therein). The beam dynamics and the 
continuity of the wave function define the quantum state of the particle after the penetration. This state cannot be a plane wave in vacuum and one of Landau states in another solenoid (except for the case of $\widetilde{\bm B}=-\bm B$). It should be an appropriate twisted beam in vacuum and a spatially periodic beam (discovered in Ref. \cite{arXiv}) in the second solenoid. In classical physics, the particle penetration to vacuum leads to a particle spread. Its analog in QM is an increase of the beam width $w(z)$. Therefore, the beam width at the boundary between the solenoid and vacuum is minimum and is equal to $w_0$. For the penetration of the Landau beam, $w_0=w_m$.

In the considered case, the quantity $\langle\mathcal{L}_z\rangle$ should be changed after the penetration despite the conservation of $\bm{\pi}^2-e\bm\Sigma\cdot\bm B$. This situation can take place only when the horizontal and vertical components of the kinetic OAM are simultaneously changed. It has been noted in Sec. \ref{Sect2} that the total momentum is turned in the plane orthogonal to $\bm e_r$. The consideration of the momentum dynamics in a real solenoid explains all related problems. In addition, this consideration does not need the above-mentioned approximation of the sharp boundary between the solenoid and vacuum or another solenoid.

To describe the quantum-mechanical dynamics at a particle beam penetration, we use the same approach as in Sec. \ref{Sect2}. However, the dynamics of a charged point considered in Sec. \ref{Sect2} differs from that of a quantum object. We use here the gauge (\ref{eqelfll}). The particle dynamics is described by Eq. (\ref{eqdynal}). The semiclassical equation (\ref{emeqp}) is also applicable. The quantities $r$ and $R_0$ influence the intrinsic and extrinsic OAMs, respectively. The intrinsic OAM characterizes quantum-mechanical properties of a beam and the extrinsic one defines the motion of the beam as a whole. 

When the particle penetrates into vacuum or any antiparallel magnetic field $\widetilde{\bm B}=-|\widetilde{B}|\bm e_z$, local changes of the intrinsic kinetic momentum of the beam are given by 
Eq. (\ref{eqinm}).
These changes do not lead to the beam motion as a whole. 

The corresponding change of the extrinsic kinetic momentum of the beam reads
\begin{equation}
\Delta\pi^{(0)}_\phi=\int{\frac{eB_{R_0}\pi_z}{\epsilon}dt}=-\frac{eR_0}{2}\int{\frac{\partial B_z}{\partial z}dz}=\frac{eR_0(B+|\widetilde{B}|)}{2}.
\label{eqent}
\end{equation} This quantity defines the motion of the beam as a whole.

Of course, the fields of the solenoids can be parallel. In this case, $\widetilde{\bm B}=|\widetilde{B}|\bm e_z$ and the change $B+|\widetilde{B}|\rightarrow B-|\widetilde{B}|$ should be made in Eqs. (\ref{eqinm}) and (\ref{eqent}). 
For vacuum, $\widetilde{B}=0$. In any case, the penetration does not change quantum numbers of the beam.

After the penetration to the vacuum, the additional transversal kinetic momentum $\Delta\pi^{(0)}_\phi$ is the same for any partial beam. The intrinsic OAM of the particle is always conserved because this momentum is acquired by the particle as a whole. The total OAM is also conserved because the contributions of $\bm A$ and of the additional transversal momentum are equal to each other. 

In the case of vacuum, the penetrated beam can come from different parts of the solenoid and its motion as a whole can result in a significant particle momentum spread. This effect can be undesirable. We underline that such a spread of \emph{momenta} of twisted beams should not be confused with the spread of twisted states. Any particle in a twisted state moves as a whole and is a stable object which motion is non-spreading (see Refs. \cite{FloettmannKarlovets,Karlovets2021} and our precedent explanation).
The particle momentum spread vanishes when the axes of symmetry of the quantum state and the solenoid coincide. Therefore, the experimental conditions proposed in Refs. \cite{FloettmannKarlovets,Karlovets2021} are optimal only on condition that $\bm \pi_\bot(t=0)=0$. If $\bm \pi_\bot(t=0)$ is nonzero, a target containing produced particles should be displaced relative to the solenoid axis by the distance $R=|\bm \pi_\bot(t=0)|/|eB|$.

In vacuum, extrinsic canonical and kinetic OAMs are equal to zero.

It is instructive to analyze the more general case when the axes of symmetry of two solenoids are parallel but do not coincide.
%
Even in the simplest case of coinciding magnetic fields in the solenoids ($\bm{B}=\widetilde{\bm B}$), there is the new effect of a change of the \emph{intrinsic} OAM after the penetration to the second solenoid. In this case, the radial fields in the two solenoids are not collinear and are given by
\begin{eqnarray}
\bm B_R=-\frac{R}{2}\frac{\partial B_z}{\partial z}\bm e_R,\qquad \bm{\widetilde{B}}_{\widetilde{R}}=-\frac{\widetilde{R}}{2}\frac{\partial \widetilde{B}_z}{\partial z}\bm e_{\widetilde{R}},
\qquad\bm R=\bm R_0+\bm r,\qquad \widetilde{\bm R}=\widetilde{\bm R}_0+\bm r.
\label{eqelflg}\end{eqnarray}
Equation (\ref{eqelflg}) shows that the radial magnetic field defined by Eq. (\ref{eqelfillll}) does not influence the vector potentials of solenoids. If the term proportional to $B''_z$ can be neglected and the distance between the solenoid axes is defined by the vector $\bm d$ (see Fig. 1), we obtain $\bm B(\bm R_0)=\widetilde{\bm B}(\widetilde{\bm R}_0)=\bm B$ and $\widetilde{\bm R}=\bm R-\bm d$.

The particle penetration to the second solenoid leads to the appearance of the transverse component of the kinetic momentum of the beam as a whole, $\Delta\bm\pi_\bot^{(0)}$, 
and to a turn of the total particle momentum in the vertical plane orthogonal to $\Delta\bm\pi_\bot^{(0)}$. Here $\Delta\bm\pi_\bot=0$. As follows from Eq. (\ref{eqdynll}) and the precedent explanations,
\begin{equation}
\Delta\bm\pi_\bot^{(0)}=\frac{e(\bm B-\bm B^{(0)})\times\bm R}{2}-\frac{e(\bm B-\bm B^{(0)})\times\widetilde{\bm R}}{2}=
\frac{e(\bm B-\bm B^{(0)})\times\bm d}{2},
\label{eqelfgg}\end{equation}
where $\bm B^{(0)}$ is the magnetic field between the solenoids. It 
can be properly calculated because the distribution of currents is always well defined.

The horizontal motion of the beam as a whole in the second solenoid leads to an extra orbital motion changing the kinetic and canonical OAMs of the twisted particle in this solenoid. We suppose that this motion is classical and underline that the magnetic field remains unchanged. Evidently, $\Delta\bm\pi_\bot=0$ and the kinetic OAM existing in the first solenoid conserves after the penetration. The new \emph{intrinsic} kinetic OAM additionally includes the term 
$$\boldsymbol{\mathcal{R}}\times\Delta\bm\pi_\bot^{(0)},$$
where $\boldsymbol{\mathcal{R}}$ is the classical radius vector of the orbit corresponding to the extra orbital motion and defined by Eq. (\ref{OAMcl}).

Thus, two new wonderful effects take place in the considered case. The first effect is a change of the intrinsic OAM of the twisted particle after the penetration to another solenoid with the same magnetic field and the shifted symmetry axis (see Fig. 1). The second effect consists in a shift of the axis of symmetry of the quantum state because of an additional beam motion with the kinetic momentum $\Delta\bm\pi_\bot^{(0)}$.

We can now pass to the rather general case of the penetration, when the second solenoid, in addition to shifting the symmetry axis, has the different magnetic field $\widetilde{\bm B}\neq\bm B$. In this case, the total change of the particle momentum is determined similarly to Eqs. (\ref{eqinm}), (\ref{eqent}), and (\ref{eqelfgg}). The beam dynamics in the two solenoids is described by two integrals which generalize the integrals used in Eq. (\ref{eqelfgg}). Their calculation results in
\begin{equation}
\begin{array}{c}
\Delta\bm\pi_\bot^{(total)}=\frac{e(\bm B-\bm B^{(0)})\times\bm R}{2}-\frac{e(\widetilde{\bm B}-\bm B^{(0)})\times\widetilde{\bm R}}{2}=
\frac{e(\bm B-\widetilde{\bm B})\times(\bm R_0+\bm r)}{2}+\frac{e(\widetilde{\bm B}-\bm B^{(0)})\times\bm d}{2},\\
\Delta\bm\pi_\bot=\frac{e(\bm B-\widetilde{\bm B})\times\bm r}{2},\qquad \Delta\bm\pi_\bot^{(0)}=\frac{e(\bm B-\widetilde{\bm B})\times\bm R_0}{2}+\frac{e(\widetilde{\bm B}-\bm B^{(0)})\times\bm d}{2}.
\end{array}
\label{eqelfnw}\end{equation}

The internal canonical momentum and, therefore, the intrinsic canonical OAM are conserved immediately after the penetration [see Eq. (\ref{eninm})]. However, there is also the additional external canonical momentum $\Delta\bm p_\bot$. It is converted into the internal one because the twisted particle rotates as a whole in the magnetic field $\widetilde{\bm B}$. 

The corresponding part of the intrinsic kinetic OAM is given by
\begin{equation}
\langle\widetilde{\mathcal{L}}_z\rangle=\langle
\mathcal{L}_z\rangle+\frac{e}{2}(B_z-\widetilde{B}_z)\langle r^2\rangle=\hbar\ell-\frac{e}{2}\widetilde{B}_z\langle r^2\rangle.
\label{radkOAM}
\end{equation}

The rest of $\Delta\bm\pi^{(total)}_\bot$ defining the second contribution reads
$$\Delta\bm\pi^{(0)}_\bot=\frac{e}{2}\left(\bm B\times\bm R_0-\widetilde{\bm B}\times\widetilde{\bm R}_0\right)=\frac{e}{2}\left(\bm B-\widetilde{\bm B}\right)\times\bm R_0+\frac e2(\bm B-\bm B^{(0)})\times\bm d.$$

Certainly, the vector $\Delta\bm\pi^{(0)}_\bot$ rotates in the second solenoid. 

According to Eq. (\ref{OAMcl}), the classical radius of the corresponding circular orbit is equal to
\begin{equation}
\mathfrak{R}=\frac{|\Delta\bm\pi^{(0)}_\bot|}{|e\widetilde{\bm B}|}.
\label{radcl}
\end{equation} The radius vector of the orbit $\boldsymbol{\mathcal{R}}$ is orthogonal to $\Delta\bm\pi^{(0)}_\bot$ and $\widetilde{\bm B}$.
The direction of rotation is unambiguously defined by the Lorentz force. The corresponding kinetic OAM $\Delta\boldsymbol{\mathcal{L}}^{(0)}$ should satisfy the relation ${\rm sgn}(e\widetilde{B}_z)\Delta\mathcal{\bm L}^{(0)}_z\le0$. The point $\widetilde{\rm C}$ denoting the center of the circle formed by the beam rotating in the second solenoid is characterized by the radius vector $\boldsymbol{\mathfrak{R}}_0$ (see Fig. 1). As a result, the axis of symmetry of the quantum state in the second solenoid is shifted at the distance $\mathfrak{R}$ as compared with the related axis in the first solenoid. The penetration changes the \emph{intrinsic} kinetic and canonical OAMs. As follows from Eqs. (\ref{numberd}) and (\ref{OAMcl}), the classical formula for the contributions caused by $\Delta\bm\pi^{(0)}_\bot$ is given by
\begin{equation}
\langle\Delta\mathcal{L}^{(0)}_z\rangle=2\Delta L^{(0)}_z=-e\widetilde{B}_z\mathfrak{R}^2.
\label{radcOAM}
\end{equation}
The corresponding quantum-mechanical formula reads
\begin{equation}
\Delta L^{(0)}_z=\hbar\mathfrak{l},\qquad 
\langle\Delta\mathcal{L}^{(0)}_z\rangle=\Delta L^{(0)}_z-\frac{e}{2}\widetilde{B}_z\mathfrak{R}^2,
\label{radqOAM}
\end{equation} where $\mathfrak{l}$ is integer. When $\mathfrak{l}\gg1$, $\hbar\mathfrak{l}$ should be close to $-(e/2)\widetilde{B}_z\mathfrak{R}^2$. The total intrinsic kinetic and canonical OAMs are given by
\begin{equation}
\widetilde{L}^{(i)}_z=\hbar(\ell+\mathfrak{l}),\qquad \langle\widetilde{\mathcal{L}}^{(i)}_z\rangle=\widetilde{L}^{(i)}_z-e\widetilde{B}_z\langle(\boldsymbol{\mathfrak{R}}+\bm r)^2\rangle=\widetilde{L}^{(i)}_z-e\widetilde{B}_z(\mathfrak{R}^2+r^2).
\label{radtOAM}
\end{equation}
We underline that QM does not admit a detachment of the canonical OAMs $\hbar\ell$ and $\hbar\mathfrak{l}$ when the beam is observed far from the boundary. Only the total intrinsic canonical OAM $\widetilde{L}^{(i)}_z$ can be defined.

We can conclude that the complete quantum-mechanical picture of the penetration is very nontrivial due to the effect described in Refs. \cite{FloettmannKarlovets,Karlovets2021}. The intrinsic canonical and kinetic OAMs in the first solenoid are defined by Eqs. (\ref{numberd}) and (\ref{ckOAM}). The intrinsic canonical OAM $L_z$ existing in the first solenoid remains unchanged in the second solenoid, but it is added by the quantity $\Delta L^{(0)}_z$. This effect is rather nontrivial because even the direction of $\widetilde{L}_z$ is unnatural when the field directions in the solenoids are antiparallel. We have shown that the penetration also leads to the appearance of the additional beam rotation and shifting the axis of symmetry of the quantum state in the second solenoid.

\begin{center}
\begin{figure}
\includegraphics[width=0.5\textwidth]{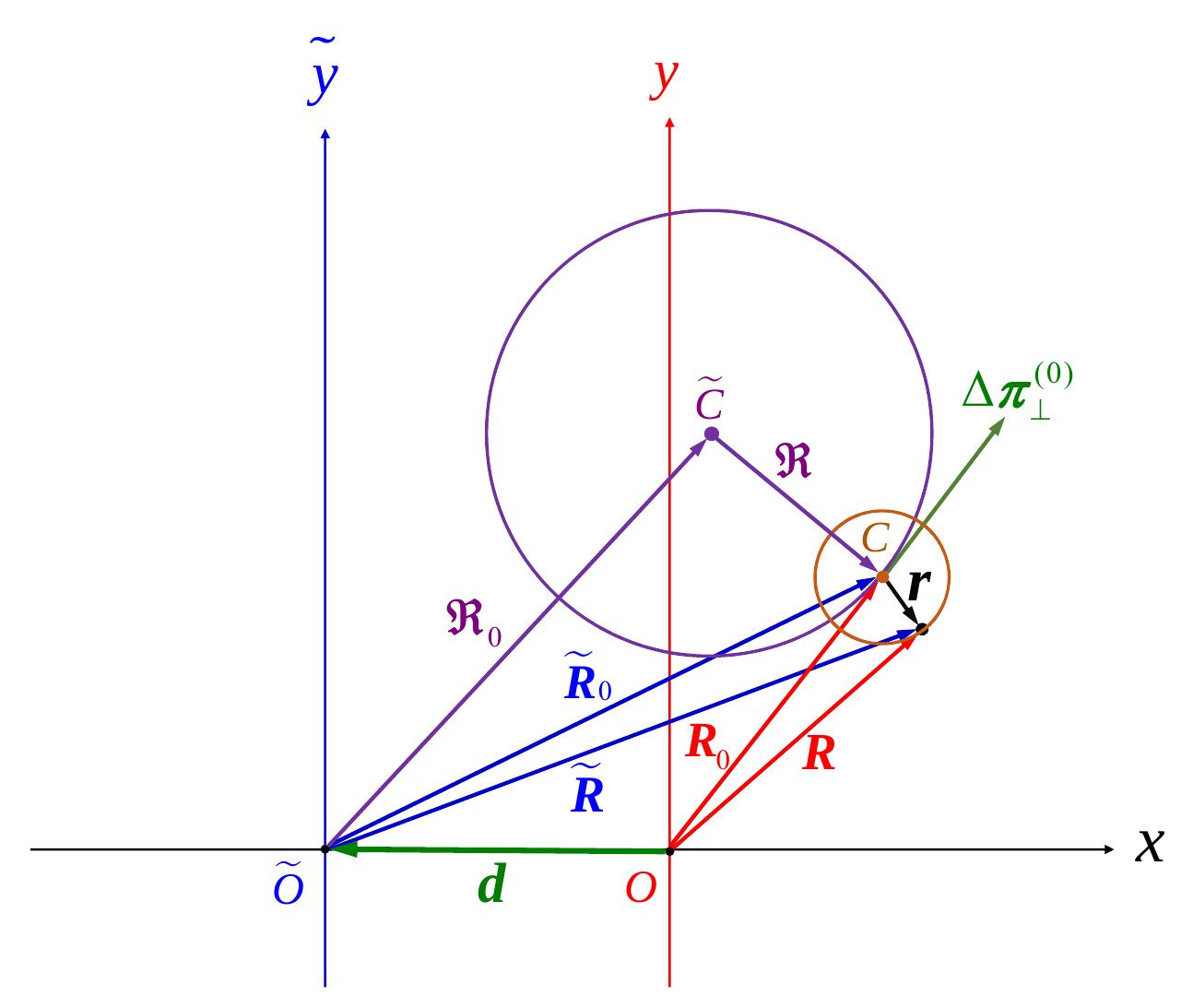}
\caption{Particle penetration between two solenoids with parallel but shifted axes of symmetry. In the second solenoid, the particle beam moves as a whole with the kinetic momentum $\Delta\bm\pi_\bot$.}
\label{fig1}
\end{figure}
\end{center}

\section{Evolution of quantum states of charged particles at the penetration through a solenoid}\label{QntSt}

In the precedent section, we mostly used the semiclassical description of the particle beam penetration to the vacuum or another solenoid. 
The semiclassical method is sufficient for an exhaustive description of the particle beam penetration to the vacuum. However, the particle beam penetration to another solenoid is substantially different and needs a quantum-mechanical analysis. After the penetration to another solenoid, unlike that to the vacuum, the \emph{intrinsic} OAM of the beam is changed. 

The problem of the final OAM of the beam with the zero initial OAM which passes through a solenoid has been briefly discussed in Ref. \cite{FloettmannKarlovets}. It has been noted that if the beam enters the solenoid with the zero OAM, it exits also with the zero OAM. However, only the total OAM conserves and the intrinsic OAM can be changed. Therefore, the problem of conservation of the intrinsic OAM after the beam passage through the solenoid needs a careful quantum-mechanical analysis. To simplify the analysis, we can use the approximation of sharp boundaries between the solenoid and vacuum. In this case, the time of flight of a particle inside the solenoid is defined by the longitudinal particle momentum. When the initial particle state is a de Broglie wave, this time is fixed. It becomes unfixed if the initial particle state is multiwave and both the longitudinal and transversal particle momenta vary for different partial waves. When the time of flight inside the solenoid is fixed, the circular motion of the particle in the transversal plane can be characterized by the \emph{definite} phase $\phi$ only when the angular velocity of this motion $\omega$ commutes with the Hamiltonian:
$$\omega\equiv\frac{d\phi}{dt}=\frac i\hbar\left[{\cal H}_{FW},\phi\right],\qquad \left[{\cal H}_{FW},\omega\right]=0.$$

However, the latter relation does not satisfied in the considered case. The exact FW Hamiltonian has the form (\ref{eq12new}) and the operator $\omega$ is defined by

\begin{equation}
\omega=\frac{1}{2}\left\{\frac{\pi_\phi}{r},\frac{1}{\sqrt{m^2+\bm{\pi}^2-e\bm\Sigma\cdot\bm B}}\right\},\qquad  \frac{\pi_\phi}{r}=\frac{p_\phi}{r}-\frac{eB}{2}=-\frac{i\hbar}{r^2}\frac{\partial}{\partial\phi}-\frac{eB}{2}.
\label{omeganc}
\end{equation} We consider positive-energy states and omit the matrix $\beta$.
Evidently, the operator $r$ does not commute with the FW Hamiltonian and $[{\cal H}_{FW},\omega]\neq0$. Specifically, 
$$ \left[\bm\pi^2,\frac{1}{r^2}\right]=-i\hbar\left[\bm\pi\cdot\nabla\left(\frac{1}{r^2}\right)
+\nabla\left(\frac{1}{r^2}\right)\cdot\bm\pi\right]=2i\left\{p_r,\frac{1}{r^3}\right\},$$
$$ i\left[{\cal H}_{FW},\frac{1}{r^2}\right]=-\frac12\left\{\frac{1}{\sqrt{m^2+\bm{\pi}^2-e\bm\Sigma\cdot\bm B}},\left\{p_r,\frac{1}{r^3}\right\}\right\},$$
and
\begin{equation}
\frac{d\omega}{dt}=\frac i\hbar\left[{\cal H}_{FW},\omega\right]=-\frac14\left\{\frac{1}{\sqrt{m^2+\bm{\pi}^2-e\bm\Sigma\cdot\bm B}},
\left\{\frac{L_z}{\sqrt{m^2+\bm{\pi}^2-e\bm\Sigma\cdot\bm B}},\left\{p_r,\frac{1}{r^3}\right\}\right\}\right\},\qquad  L_z=-i\frac{\partial}{\partial\phi}.
\label{omegafn}
\end{equation}
The operator $L_z$ commutes with the FW Hamiltonian. Certainly, $$\left\{p_r,\frac{1}{r^3}\right\}\Psi_{FW}=-i\frac{2}{r^3}\frac{\partial\Psi_{FW}}{\partial r}+i\frac{3\Psi_{FW}}{r^4}.$$

The phase defined by the direction of $\bm\pi_\bot$ remains unchanged for the beam in the vacuum but becomes uncertain in a magnetic field.

For particles entering the solenoid in any multiwave state, the additional factor of noncommutation of ${\cal H}_{FW}$ and $\omega$ is a difference of longitudinal momenta of different partial waves. As a result, the particle penetration into the solenoid leads to a uncertainty of phase of the particle rotation. This uncertainty increases with the distance passed by the particle. When the solenoid length is small, the particle exits as a de Broglie wave. When this length is large, all final phases become equally probable and the particle exits as a Laguerre-Gauss beam. 

Evidently, the intrinsic OAM $\widetilde{L}^{(i)}_z$ defined by Eq. (\ref{radtOAM}) does not depend on a phase. Therefore, it is not changed during the rotation and the phase uncertainty is followed by the stable conserved OAM.

A similar situation takes place when the charged particle penetrating through the solenoid is in a multiwave state. In such states, there is the additional factor of the difference of longitudinal momenta of different partial waves accelerating the appearance of the uncertainty of phase of the particle rotation. When the solenoid length is rather small, this uncertainty can be neglected and the particle exits the solenoid with the same \emph{intrinsic} OAM, $L_z^{(i)}$, as in the initial state. When this length is large enough, the \emph{intrinsic} OAM of the exiting particle, $\widetilde{L}^{(i)}_z$, significantly differs from the \emph{intrinsic} OAM of the entering particle. This effect can open exciting new possibilities of production of twisted particles in real solenoids without additional devices like particle sources. However, it still needs an experimental confirmation.

\section{Production of charged twisted particles and a measurement of their orbital angular momenta in magnetic fields}\label{Sect5}

The fulfilled theoretical analysis allows us to  propose the further development of the method of production of charged twisted particles in magnetic fields proposed in Refs. \cite{FloettmannKarlovets,Karlovets2021}. Our development allows one to discover twisted positrons and positroniums as well as some other charged twisted particles. It is also based on the quantum Busch theorem (see Refs. \cite{FloettmannKarlovets,Karlovets2021}) but uses a specific experimental design. We propose to utilize a standard positron source placed in a quasi-uniform magnetic field of a real solenoid. The results presented in Sec. \ref{Sect2} show that charged particles produced in uniform and even nonuniform magnetic fields have the definite direction of the OAMs. Their orbital polarization can be rather large. In this case, a spin polarization of produced particles can be neglected. The connection between the canonical and kinetic OAMs of created particles is analyzed in Sec. \ref{Sect3}. A variation of the direction and value of the magnetic induction in the solenoid changes the direction and values of the OAMs. 

We suppose that emitted positrons or other particles penetrate from the solenoid 
and then are governed by a quadrupole magnetic field. This field acts on large magnetic moments of twisted particles caused by their large OAMs \cite{Barut}:
\begin{equation}
\begin{array}{c}
\bm\mu=\frac{e(\bm L+2\bm s)}{2\gamma m},
\end{array}
\label{mdm}
\end{equation} where $\gamma$ is the Lorentz factor. As a result, the particle motion is not rectilinear. A narrow source outlet decreases the beam momentum spread after the penetration. 
Let the $z$ axis of the Cartesian coordinate system coincides with the symmetry axes of the solenoid and source of particles. Similarly to the Stern-Gerlach experiment \cite{SternGerlach}, one should differently deflect particles with different orbital polarizations. Since OAMs of all particles are parallel and antiparallel to the $z$ axis, one needs to use a magnet with maximum $|\widetilde{B}_z|$, minimum $|\widetilde{B}_x|$, and $\widetilde{B}_y=0$. In this case, 
\begin{equation}
\begin{array}{c}
\frac{\partial \widetilde{B}_z}{\partial x}=\frac{\partial \widetilde{B}_x}{\partial z},\qquad \widetilde{B}_x=\frac{\partial \widetilde{B}_z}{\partial x}z. \end{array}
\label{qmf}
\end{equation}
We can present this magnetic field as the sum of the uniform field $\widetilde{\bm B}=\widetilde{B}_z(x=y=0)\bm{e}_z\equiv\widetilde{B}\bm{e}_z$ and the quadrupole magnetic field $$\widetilde{B}_x=\kappa z,\qquad\widetilde{B}_y=0,\qquad\widetilde{B}_z=\kappa x,\qquad\kappa=\frac{\partial\widetilde{B}_z}{\partial x}.$$
We suppose that $\widetilde{B}_z(x=y=0)>0$ and the magnetic field in the solenoid section, $\bm B$, can be parallel and antiparallel to the $z$ direction.

It is convenient to apply the magnetic field lines tangent to the vector $\widetilde{\bm B}$. These lines are defined by the equation $$\frac{dx}{dz}=\frac{\widetilde{B}_x}{\widetilde{B}_z}=\frac{\kappa z}{\widetilde{B}+\kappa x}.$$
The line containing the point $(x_0,y_0=0,z_0)$ and tangent to $\widetilde{\bm B}$ has the form
\begin{equation}
\begin{array}{c}
z^2=\left(x+\frac{2\widetilde{\bm B}}{\kappa}\right)x-\left(x_0+\frac{2\widetilde{\bm B}}{\kappa}\right)x_0+z_0^2.
\end{array}
\label{mline}
\end{equation} Since the inversion $z\rightarrow-z$ takes place, it is convenient to choose the points $(x_0,y_0=0,z_0=0)$ and vary $x_0$. When $\kappa$ is small, the magnetic field lines are parabolas:
\begin{equation}
\begin{array}{c}
z^2=\frac{2\widetilde{\bm B}(x-x_0)}{\kappa}.
\end{array}
\label{mlinepa}
\end{equation}

Evidently, needed magnet parameters in the proposed experiment and the Stern-Gerlach one \cite{SternGerlach} substantially differ. The force acting on twisted particles is given by
\begin{equation}
\begin{array}{c}
\bm f=\nabla(\bm\mu\cdot\widetilde{\bm B}),\qquad f_x=\frac{eL_z}{2\gamma m}\frac{\partial \widetilde{B}_z}{\partial x}.
\end{array}
\label{force}
\end{equation}
It deflects twisted particles in the $x$ direction. Importantly, the Lorentz force acting due to the weak field $\widetilde{B}_x$ on \emph{charges} leads to a deflection in the $y$ direction. The deflection is independent of $L_z$.

It is very important that twisted particles created in the uniform magnetic field remains twisted and non-spreading 
after the penetration to vacuum \cite{FloettmannKarlovets,Karlovets2021} or to a different magnetic field (see Sec. \ref{Sect4}). In the latter case, the beam width spatially oscillates \cite{arXiv}. When the axes of symmetry of the solenoid and a quantum state of a twisted particle coincide, the particle penetrating into vacuum or a different uniform magnetic field moves in the same direction (the $z$ axis). In a nonuniform magnetic field, the needed deflection is provided by the force (\ref{force}). When the above-mentioned axes do not coincide, the penetrating particle does not move along the solenoid axis (i.e., $x\neq0$ or $y\neq0$). Certainly, such a behavior which reason is expounded in Sec. \ref{Sect4} is rather unwanted.  As a result, the diameter of the outlet of the particle source should be small enough in comparison with the deflection of the twisted particle. This is very possible because the deflection is proportional to its OAM which can be rather large. Thus, some twisted particles can be successfully discovered.

We propose to use the developed  experimental design for a discovery of twisted positrons (see Fig. 2). We suggest to apply a solenoid with a variable (quasi)uniform field for a production of twisted positron beams which then penetrate to the quadrupole magnetic field (\ref{qmf}). Their OAMs can be measured because beams with different OAMs are registered in different parts of the target. The blue and green lines demonstrate the beams produced at \emph{opposite} field directions in the solenoid. The field direction shown on Fig. 2 corresponds to the OAM $\bm L_1$.

\begin{center}
\begin{figure}
\includegraphics[width=0.5\textwidth]{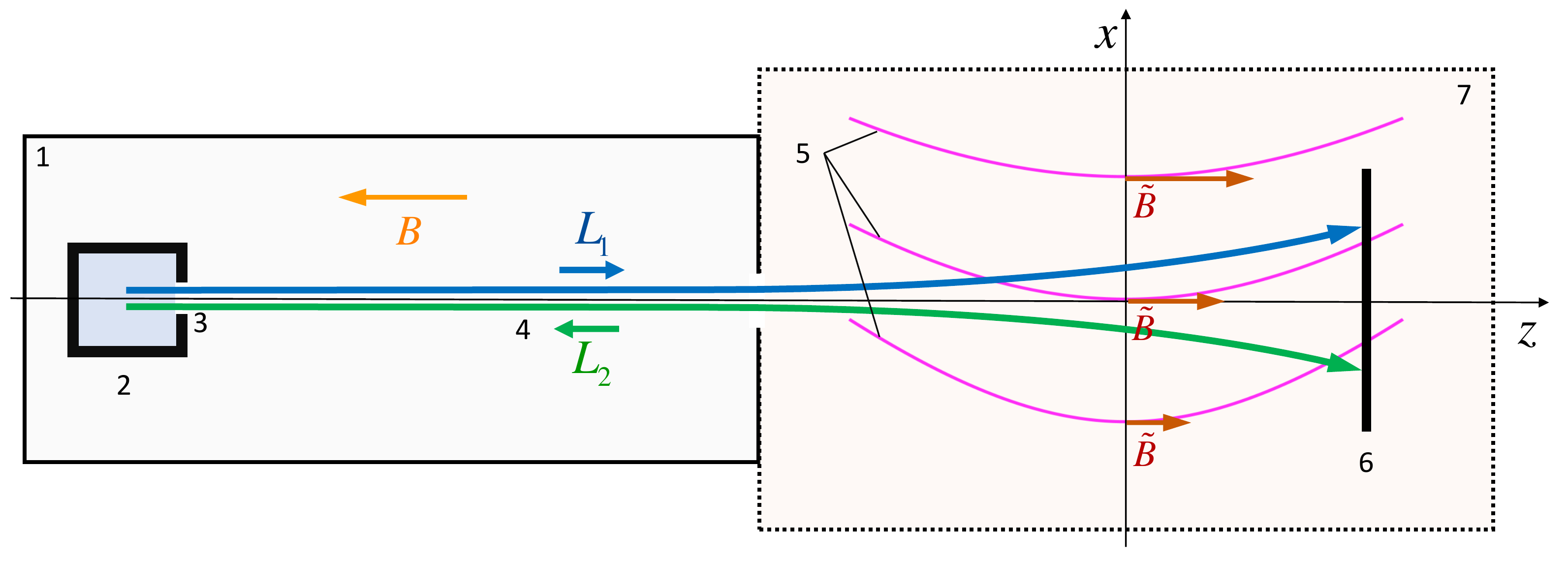}
\caption{Experimental design appropriate for a discovery of twisted positrons.  
Such positrons are produced by the source 2 placed in the solenoid 1 with the (quasi)uniform magnetic field $\bm B$. The direction and value of this field can be varied. The two twisted positron beams 4 with OAMs $\bm L_1$ and $\bm L_2$ correspond to the field direction shown in the figure and to the opposite direction, respectively. Twisted beams penetrating to the magnet 7 with the non-uniform magnetic field $\widetilde{\bm B}$ are registered by the target 6. Other denotations are: 3 is the source outlet and 5 are the field lines.}
\label{fig2}
\end{figure}
\end{center}

\section{Potential for a discovery of twisted positroniums
}\label{SecPs}

Another exiting possibility is a discovery of twisted positroniums. A positronium is an atom formed by an electron and a positron. Parapositroniums and orthopositroniums have spins 0 and 1, respectively. Twisted positrons are emitted in the solenoid in the magnetic field $\bm B$ and then penetrate to vacuum or a medium. Their effective masses in vacuum are given by \cite{photonPRA}
\begin{equation}
\begin{array}{c}
M=\sqrt{m^2+<\bm p_\bot^2>}=\sqrt{m^2+\frac{2(2n+|\ell|+1)}{w_0^2}}.
\end{array}
\label{masselv}
\end{equation}
Here $w_0=w_m$ (see Sec. \ref{Sect4}) and $w_m$ is given by Eq. (\ref{Lenergy}). When $B=1$ T, $w_m=5.1\times10^{-8}$ m and the quantities $1/w_m$ and $1/(2mw_m^2)$ are equivalent to 3.9 eV and 1.5$\times10^{-5}$ eV, respectively. When the OAM is relatively large, $|\ell|\approx10^4$, and $n\sim1$, $2(2n+|\ell|+1)/(2mw_m^2)$ is approximately equivalent to 0.3 eV.  The last value defines the difference $M-m$. 

After the penetration, the twisted positron can pick up an electron and can form a positronium which should also be twisted. Its mass $m_{Ps}$ is equal to the doubled electron mass corrected for the binding energy (approximately 6.8 eV). The effective mass of the twisted positronium in vacuum is also defined by Eq. (\ref{masselv}) (with $m_{Ps}$ instead of $m$) but $w_0$ can be different. In the frame, in which the momentum of the twisted positronium vanishes, its total energy is equal to $Mc^2$. To form the positronium, the electron and positron should have a smaller total kinetic energy in their center-of-mass frame than the binding energy. This requirement leads to a need of a thermalization (deceleration) of the positron. This goal can be achieved with an appropriate electric field in vacuum or with passing the positron throughout a medium. The momentum and OAM of the positron are usually collinear. Since the decelerating electric field is antiparallel to the positron momentum, the first method conserves the positron OAM. The second method decreases it because the positron can loose the OAM owing to interactions with the medium. When the difference $M-m_{Ps}$ is less than the
binding energy of the positronium, the spontaneous decomposition of Ps into e$^+$ and e$^-$ is impossible. Therefore, it is preferable when the condition $M-m_{Ps}<6.8$ eV is satisfied. The estimate made in the previous paragraph shows that its satisfaction always (or almost always) takes place. 

To determine the difference between the effective mass of positronium and $m_{Ps}$, one can monitor the positronium annihilation. Orthopositroniums and parapositroniums annihilate with an emission of three and two gamma quanta, respectively. The total momentum of these quanta is equal to zero in the frame in which their total energy is equal to $Mc^2$. Since measurements of positronium annihilation have a high precision, a determination of difference $M-m_{Ps}\sim0.1$ eV is quite possible.

Thus, the proposed experiment with the positron source within a solenoid allows one a successful discovery of the twisted positronium.

\section{Discussion and summary}\label{SectC}

Important results obtained by Floettmann and Karlovets \cite{FloettmannKarlovets} and Karlovets \cite{Karlovets2021} have opened wonderful possibilities for discoveries of new twisted particles having the OAMs. In the present study, we have considered these possibilities in more details and have determined a potential for discoveries of twisted positrons and positroniums. 

New possibilities found in Refs. \cite{FloettmannKarlovets,Karlovets2021} are based on the use of a (quasi)uniform magnetic field of a solenoid. We have specified quantum states suitable for a production of charged particles in a uniform magnetic field. The analysis of canonical and kinetic OAMs of particles produced in a (quasi)uniform field of real solenoids has been fulfilled. We have analyzed in detail the particle penetration from a solenoid to vacuum or another solenoid. We have shown that it usually leads to the appearance of the additional beam rotation and shifting the axis of symmetry of the quantum state in the second solenoid. Finally, we have developed experiments allowing one successful discoveries of twisted positrons and positroniums. We have proposed to place a positron source into a solenoid for producing twisted positrons and, as a result of an interaction with a medium, twisted positroniums. We have also developed methods of checking their twisting. We suppose that these methods can also be applicable to other charged particles.

In addition, we have found the new effect of increasing a uncertainty of phase of the particle rotation with the distance passed by the particle in the solenoid. This effect leads to exciting new possibilities of production of twisted particles in real solenoids without additional devices like particle sources.

\begin{acknowledgments}
The work was supported by the National Natural Science
Foundation of China (grants No. 12175320, 11975320 and No. 11805242), the Natural Science Foundation of Guangdong Province, China (grant No. 2022A1515010280), and by the Chinese Academy of Sciences President's International Fellowship Initiative (grant No. 2019VMA0019).
A. J. S. also acknowledges hospitality and support by the
Institute of Modern
Physics of the Chinese Academy of Sciences.
\end{acknowledgments}


\begin{thebibliography}{}
\bibitem{footnote}
Of course, the sharp boundary between two areas is an approximation. However, we consider below real solenoids which fields are slightly nonuniform. Our study is mostly based on classical and quantum-mechanical equations of motion and does not need the approximation of the sharp boundary between two solenoids or the solenoid and vacuum.
\end{thebibliography}
\end{document}